\documentclass[12pt,aps,prl,floats,graphicx,euscript,amsmath,amssymb,
onecolumn]{revtex4}
\usepackage[cp1251]{inputenc}
\usepackage[dvips]{graphicx}
\DeclareMathAlphabet\EuScript{U}{eus}{m}{n} \SetMathAlphabet\EuScript{bold}{U}{eus}{b}{n}

\def\lapprox{\,\raise0.4ex\hbox{$<$}\kern-0.8em\lower0.7ex\hbox{$\sim$}\,}
\def\gapprox{\,\raise0.4ex\hbox{$>$}\kern-0.8em\lower0.7ex\hbox{$\sim$}\,}
\begin{document}
\bibliographystyle{prsty}
\title{Goldstone Mode Relaxation in a Quantum Hall Ferromagnet due to Hyperfine Interaction with Nuclei}

\author{S. Dickmann}
\affiliation{Institute for Solid State Physics of RAS, Chernogolovka 142432, Moscow
District, Russia}

\begin{abstract}
\vspace{0.mm} Spin relaxation in quantum Hall ferromagnet regimes
is studied. As the initial non-equilibrium state, a coherent
deviation of the spin system from the ${\vec B}$ direction is
considered and the breakdown of this Goldstone-mode state due
to hyperfine coupling to nuclei is analyzed. The relaxation
occurring non-exponentially with time is studied in terms of
annihilation processes in the ``Goldstone condensate'' formed by
``zero spin excitons''. The relaxation rate is
calculated analytically even if the initial deviation is not small.
This relaxation channel competes with the relaxation mechanisms due to
spin-orbit coupling, and at strong magnetic fields it becomes
dominating.

\noindent PACS numbers 73.21.Fg, 73.43.Lp, 78.67.De
\end{abstract}
\maketitle

\bibliographystyle{prsty}

The reported work is aimed at description of the relaxation channel
which might become dominant at magnetic fields $B>10\,$T in the
so-called quantum Hall ferromagnet (QHF). The latter is a 2DEG
state under quantum Hall effect conditions where all electrons
of the upper, not completely filled Landau level, have in the
ground state spins aligned along ${\vec B}$.  This
spin polarization obviously arises at odd integer fillings: $\nu=1,3,...\;$
\cite{by81}. Besides, experiments and semi-phenomenological
theories show that at some fractional fillings, namely at
$\nu=1/3,1/5,...$, electrons in the ground state occupy only one
spin sublevel, and thereby the fractional QHF state is also
realized \cite{gi85,lo93,kukush06,va09,gr08}. The QHF possesses a
macroscopically large spin ${\vec S}$ oriented in the direction of
the field ${\vec B}$ due to negative $g$-factor in GaAs
structures. In the following all calculations are carried out in
the form applicable to both odd-integer filling $\nu\!=\!2k\!+\!1$
and fractional QHF. This generalization on the $\nu\!<\!1$ case is
done in compliance with the well known semi-phenomenological
description of the fractional QHF [\onlinecite{gi85,lo93}] and, in
particular, was already used in Ref. [\onlinecite{di09}].

Obviously, there are two different types of initial deviation of
the large spin ${\vec S}$ from its equilibrium position. The first
type represents the case where vector ${\vec S}$ is changed in
length but its direction is not altered. Then the QHF symmetry is
the same as in the equilibrium state. Analysis reveals that this
type of initial perturbation is microscopically described by
excitation of spin waves where each one corresponds to the spin
numbers changed by one: $ \delta S\!=\!\delta S_z\!=\!-1\;$
\cite{by81}. The second type of spin perturbation is a coherent
deviation of ${\vec S}$ as a whole from the direction ${\hat
z}\parallel {\vec B}$ without any changes in the length of ${\vec
S}$. This case means appearance of the Goldstone mode (GM) in the
QHF, and microscopically it is described by a ``Goldstone
condensate'' of ``zero spin excitons''. Every zero spin exciton
$X_{\bf 0}$ represents a change $\delta S_z\!=\!-1$ with the total
spin kept constant: $\delta S=0$. For the first type deviation,
the relaxation was studied experimentally in Refs.
[\onlinecite{zh93,fu08,ne10}] and theoretically in Refs.
[\onlinecite{di96,di99}] and [\onlinecite{di09}]. For the second
type, the GM breakdown was theoretically analyzed in the works of
Refs. [\onlinecite{di-io96}] and [\onlinecite{di04}]. (See also
Ref. [\onlinecite{foot0}].) All these theoretical studies dealt
with the relaxation channels where spin non-conservation arose
from the spin-orbit (SO) coupling of 2D electrons.

So, as in publications of Refs. [\onlinecite{di-io96}] and
[\onlinecite{di04}], now the considered initial deviation is again
of the second type, i.e. starting point is the GM state
$|i\rangle\!=\!\left({\hat S}_-\right)^{{}\!N}\!\!|0\rangle$,
where $|0\rangle$ stands for the QHF ground state, and ${\hat
S}_-\!=\! \sum_j\hat{\sigma}^{(j)}_-$ is the lowering spin
operator. [$j$ labels electrons;
$\hat{\sigma}_\pm\!=\!(\hat{\sigma}_x\!\pm\! i\hat{\sigma}_y)/2$,
where $\hat{\sigma}_{x,y,z}$ are the Pauli matrices.] However, in
this Letter I report on another relaxation channel where the spin
non-conservation is caused by the hyperfine contact coupling to
the GaAs nuclei \cite{la77,pa77,dy84,me02}:
\begin{equation}\label{hf1}
{\hat H}_{\rm hf}=\frac{8\pi}{3}\mu_B\sum_n\frac{\mu_n}{I^{(n)}}\left(\hat{\mbox{\boldmath $I$}}^{(n)}\!\!\!\!\cdot\!\hat{\mbox{\boldmath
 $\sigma$}}\right)\delta(\mbox{\boldmath
 $R$}\!-\!\mbox{\boldmath
 $R$}_n)\,.
\end{equation}

Here $\mu_B$ is the Bohr magneton, $\mbox{\boldmath $R$}$ is the
electron position, $\hat{\mbox{\boldmath $I$}}^{(n)}$,
$\mbox{\boldmath $R$}_n$ and $\mu_n$ are spin, position and
magnetic dipole moment of the $n$-th nucleus, respectively.
Compared to the SO coupling Hamiltonian, the interaction
\eqref{hf1} has one important feature: it violates the
translation invariance of the 2D electron system and therefore
leads not only to the electron spin non-conservation but also to
the non-conservation of the 2DEG momentum. From
the viewpoint of the GM breakdown, this means that the hyperfine
coupling is sufficient to provide the GM relaxation process
without any additional dissipation mechanisms. This property is at
variance with the SO interaction relaxation mechanisms. As far as
the SO coupling does not perturb the translation symmetry, the GM
relaxation needs additional perturbative interactions providing
the momentum and energy dissipation. These in fact are the
electron-phonon coupling \cite{di-io96} or interaction with the
smooth random potential \cite{di04}.

The Hamiltonian (\ref{hf1}) may be rewritten as \cite{dy84}
\begin{equation}\label{hf2}
 {\hat H}_{\rm hf}=\frac{v_0}{2}\sum_nA_n\Psi^*(\mbox{\boldmath
 $R$}_n)\left(\hat{\mbox{\boldmath $I$}}^{(n)}\!\!\!\!\cdot\!\hat{\mbox{\boldmath
 $\sigma$}}\right)\Psi(\mbox{\boldmath $R$}_n)\,,
\end{equation}
where $\Psi(\mbox{\boldmath $R$})$ is the electron envelope
function, and  $v_0$ is the unit cell volume. Both Ga and As
nuclei have the same total spin: $I^{\rm Ga}\!=\!I^{\rm
As}\!=\!3/2$. The parameter $A_n$, being inversely proportional to
$v_0$, actually depends only on the Ga/As nucleus position within
the unit cell. For the final calculation I only need the sum
$A_{\rm Ga}^2\!+\!A_{\rm As}^2$.  If $v_0$ is the volume of the
two atom unit cell, then using values of magnetic moments of the
Ga and As stable isotopes \cite{ra89}, I estimate that $A_{\rm
Ga}^2\!+\!A_{\rm As}^2\!\approx\!4\cdot 10^{-3}\,$meV${}^2$.

In order to describe the QHF states, I use again the {\it
excitonic representation} by analogy with previous works
\cite{di96,di-io96,di99,di04,di09}. Namely, by defining the spin
exciton creation operator \cite{dz83}:
\begin{equation}\label{Q}
  {\cal Q}_{{\bf q}}^{\dag}=\frac{1}{\sqrt{ N_{\phi}}}\sum_{p}\,
  e^{-iq_x p}
  b_{p+\frac{q_y}{2}}^{\dag}\,a_{p-\frac{q_y}{2}}\,,
\end{equation}
where $a_p$ and $b_p$ are the Fermi annihilation operators
corresponding to electron states on the upper Landau level with spin up
($a\!=\uparrow$) and spin down ($b\!=\downarrow$),
respectively ($N_\phi$ is the Landau level degeneration number), I consider
states $|N\rangle\!=\!\left(\!{\cal Q}_{{\bf
0}}^{\dag}\!\right)^N\!\!|0\rangle$ and $|N\!-\!1;1;{\bf
q}\rangle\!=\!{\cal Q}_{{\bf q}}^{\dag}\left(\!{\cal Q}_{{\bf
0}}^{\dag}\!\right)^{N\!-\!1}\!|0\rangle,\:$ where $|{\rm
0}\rangle\!=
\!|\overbrace{\uparrow,\uparrow,...\uparrow}^{\displaystyle{\vspace{-15mm}
\mbox{\tiny{$\;{N}_\phi$}}}}\,\rangle$ is the ground state at
odd-integer filling $\nu$. Both states correspond to the spin
$z$-component $ S_z\!=N_\phi/2\!-\!N$, whereas the total spin
number is $S=N_\phi/2$ and $S\!=N_\phi/2\!-\!1$, respectively.
[Index $p$ labels intrinsic Landau level states with wave functions
$\psi_p({\bf r})\!=\!(2\pi {\cal
N}_\phi)^{-1/4}e^{ipy}\varphi_k(p\!+\!x)$ in the Landau gauge,
$\varphi_k(x)$ is the oscillator function;  in Eq. (\ref{Q}) and
everywhere below we measure lengths in units of $l_B$ and wave
vectors in units of $1/l_B$.] The major advantage of these
excitonic states is that they are eigen states of the QHF at
odd-integer $\nu\,$:
\begin{equation}\label{eigen}
\begin{array}{l}
\quad\left[\epsilon_{\rm Z}{\hat S}_z\!+\!{\hat H}_{\rm int},\left(\!{\cal Q}_{{\bf 0}}^{\dag}\!\right)^{N}\right]{}\!{}{}|0\rangle\!=\!N\epsilon_{\rm Z}|N\rangle,\\
\left[\epsilon_{\rm Z}{\hat S}_z\!+\!{\hat H}_{\rm int},{\cal Q}_{{\bf
q}}^{\dag}\left(\!{\cal Q}_{{\bf 0}}^{\dag}\!\right)^{N\!-\!1}\right]\!|0\rangle\!=\!(N\epsilon_{\rm Z}\!+\!{\cal E}_q)|N\!-\!1;1;{\bf q}\rangle,
\end{array}
\end{equation}
where $\epsilon_{\rm Z}\!=\!g\mu_BB$ is the cyclotron gap, ${\hat
H}_{\rm int}$ is the 2DEG Coulomb interaction Hamiltonian, and
${\cal E}_q$ is the Coulomb correlation energy of the spin exciton
having momentum ${\bf q}\,$ \cite{by81}. These equations are
accurate to the first order in parameter $r_{\rm c}\!=\!(\alpha
e^2/\kappa l_B)/\hbar\omega_c$ considered to be small ($\omega_c$
is the cyclotron frequency, $\alpha\!<\!1$ is the averaged
form-factor arising  due to finiteness of the 2D layer thickness,
$\kappa$ is the dielectric constant). Only small $q$ vectors are
relevant to the studied problem (i.e. $ql_B\!\ll\!1$ in common
units), and therefore the quadratic
approximation for the spin exciton spectrum ${\cal
E}_q\!\approx\!q^2/2M_{\rm x}$ is sufficient. The exciton
mass $M_{\rm x}$ is calculated by using the finite thickness
form-factor \cite{by81,di99,di09}, although recently $M_{\rm
x}$ was measured experimentally \cite{kukush06,ga08,kukush09}.

Now I express the hyperfine coupling Hamiltonian \eqref{hf2} in terms of the
excitonic representation. By omitting the ${\hat I}_z{\hat
\sigma}_z$ term due to its irrelevance to any spin-flip process
\cite{foot1}, and substituting into Eq. \eqref{hf2} the
Schr\"odinger operators ${\hat \Psi}^\dag(\mbox{\boldmath
$R$})\!=\!\chi(z)\sum_p\left(a^\dag_p\!+\!b^\dag_p\right)\psi_p^*(
{\bf r})$ and ${\hat \Psi}(\mbox{\boldmath $R$})\!=\!\left({\hat
\Psi}^\dag\right)^\dag$  instead of ${ \Psi}^*$ and ${ \Psi}\;$
[$\chi(z)$ is size-quantized wave function, ${\bf r}\!=\!(x,y)$],
one finds
\begin{equation}\label{hf3}
{\hat H}_{\rm hf}=\frac{v_0}{4\pi l_B^2\sqrt{{\cal N}_\phi}}\sum_{\bf q}f(q){\cal Q}_{\bf
q}\sum_nA_n|\chi(Z_n)|^2e^{i{\bf q}{\scriptsize\mbox{\boldmath $R$}}_n}{\hat
I}^{(n)}_-\quad+\quad\mbox{H.c.,}
\end{equation}
where $f\!=\!e^{-q^2/4}[L_k(q^2/2)]$ if $\nu\!=\!2k\!+\!1$ ($L_k$
is the Laguerre polynomial), and $f\!=\!e^{-q^2/4}$ if
$\nu\!<\!1$.

A set of $I_z$ spin numbers $\{M\}\!=\!(M_1,M_2,...M_n,...)$,
where every $M_n$ can assume one of the values
$-3/2,-1/2,1/2,3/2$, completely determines the state of the
nuclear system. The case where 2DEG electrons are in the state
$|N\rangle$ or $|N\!-\!1;1;{\bf q}\rangle$ and nuclei in the state
$\{M\}$ I symbolize as $\left|\{M\};N\right\rangle$ and
$|\{M\};N\!-\!1;1;{\bf q}\rangle$, respectively. Application of
the decreasing/increasing operator ${\hat I}^{(n)}_\mp$ to the
former yields
%\begin{equation}\label{action}
 ${\hat I}^{(n)}_\mp\left|\{M\},N\right\rangle=\sqrt{\left(\frac{5}{2}\!\mp\!
 M_n\right)\left(\frac{3}{2}\!\pm\! M_n\right)}\left|\{M\}^\mp_n,N\right\rangle,$
%\end{equation}
where $\{M\}^\mp_n=(M_1,M_2,...M_n\!\mp\!1,...)$.

From this point the study of the relaxation rate becomes similar
to that in Ref. [\onlinecite{di04}]. The only appreciable
difference is the presence of the nuclear component. The
temperature is again assumed to be negligible. Being of the same
order or even smaller than the uncertainty value determined by the
external smooth disorder field [\onlinecite{di04}] it is, in
particular, well smaller than the Zeeman gap $\epsilon_{\rm Z}$.
The initial state $|i\rangle$  is thus the Goldstone condensate
containing $N$ zero spin excitons:
${}\!|i\rangle\!=\!|\{M\};N\rangle$. This state is electronically
and nuclearly degenerate. The GM breakdown is studied in terms of
the transitions governed by the Fermi Golden Rule probability:
$w_{if}=(2\pi/\hbar)|{\cal M}_{if}|^2\delta(E_f-E_i)$, where the
final state $|f\rangle$ is the state where a part of the Zeeman
energy has been converted into the nonzero spin exciton kinetic
energy ${\cal E}_q$. Such a transition is the
$2X_0\!+\!\{M\}\!\!\to\!\!X_{{\bf q}^*}\!+\!\{M\}^-_n$ process, if
calculated in the lowest order of the perturbation theory. The
final state for this transition is
$|f\rangle\!=\!|\{M\}^-_n;N\!-\!2;1;{\bf q}^*\rangle$, where $q^*$
is determined by the energy conservation equation
$2\epsilon_Z\!=\!\epsilon_Z\!+\!{\cal E}(q^*)$, i.e.
$q^*\!=\!\!\sqrt{2M_{\mbox{\scriptsize x}}\epsilon_Z}$. When
calculating the transition matrix element ${\cal M}_{if}(n,{\bf
q}^*)\!=\!\langle f|{\hat H}_{hf}|i\rangle$, one may take into
account that $q^*\ll 1$. So, the squared value is
\begin{equation}\label{matre}
{}\!|{\cal
M}_{if}(n,{\bf q})|^2=\frac{v_0^2A_n^2|\chi(Z_n)|^4}{(4\pi)^2 l_B^4{\cal N}_\phi} \left(\frac{5}{2}\!-\!
 M_n\right)\!\left(\frac{3}{2}\!+\! M_n\right)\!\frac{\left|\langle{\bf q};1;N\!-\!2|{\cal Q}_{-{\bf q}}|N\rangle\right|^2}{R(N)R(N\!-\!2;1;{\bf q})}\,,
\end{equation}
where the notation $R(...)$ stands for the norm of the state
$|...\rangle$. Now, at variance with the cited works
\cite{di-io96,di04}, the expectations entering Eq. \eqref{matre}
should be calculated not only for the integer QHF but for the
fractional QHF too. The latter can be obtained within the
so-called ``single-mode approximation'' \cite{gi85,lo93}, namely:
\begin{equation}\label{expectations}
\begin{array}{l}
\langle{\bf q};1;N\!-\!2|{\cal Q}_{-{\bf q}}|N\rangle=
-\displaystyle{\frac{\nu\,'{} N!(\nu\,'{}\!N_\phi\!-\!2)!}{N_\phi^{N\!-\!1}(\nu\,'{}\!N_\phi\!-\!N)!}}
\left[1+O(q^4)\right],\\
R(N)=
\displaystyle{\frac{N!(\nu\,'{}\!N_\phi)!}{N_\phi^{N}(\nu\,'{}\!N_\phi\!-\!N)!}},\quad \mbox{and}\quad
R(N;1;{\bf q})=\displaystyle{\frac{\nu\,'{}N!(\nu\,'{}\!N_\phi\!-\!2)!}{N_\phi^{N}
(\nu\,'{}\!N_\phi\!-\!N\!-\!2)!}}\left[1+O(q^4)\right].
\end{array}
\end{equation}
Here $\nu\,{}'\!\!=\!\nu$ if $\nu\!<\!1$ or $\nu\,{}'\!\!=\!1$ if
$\nu\!=\!2k\!+\!1$ ($k=0,1,2,...$). Formulas \eqref{expectations}
are exact for odd-integer $\nu\;$
\cite{dz91} (then the $\sim\!O(q^4)$ terms vanish), but for $\nu\!<\!1$ they represent a result of the semi-phenomenological approach where the expectations are expressed in terms of the two-particle correlation function
calculated for Laughlin's states \cite{gi84}.

Using Eqs. (\ref{matre}$\,$-\ref{expectations}), and assuming that
the nuclei are unpolarized, I get the rate of the considered
$S_z\to S_z\!+\!1$ process:
\begin{equation}\label{rate}
-dN/dt=\frac{2\pi}{\hbar}\sum_{n,{\bf q}}
  \left|{\cal M}_{if}(n,{\bf q})\right|^2\delta
  (q^2l_B^2/2M_{\mbox{\scriptsize  x}}\!-\!\epsilon_Z)\!=\!
  \displaystyle{\frac{N(N\!-\!1)}{\nu\,{}'N_{\phi}\tau_{\rm hf}}} \quad\mbox{(for {\it any} $N\ge 1$)},
\end{equation}
where
\begin{equation}\label{hftime}
1/\tau_{\rm hf}=\displaystyle{\frac{5v_0M_{\rm x}\left(A_{\rm Ga}^2\!+\!A_{\rm
As}^2\right)}{8\hbar l_B^2d}}\,.
\end{equation}
Here $d$ stands for a conventional width of the 2DEG:
$1/d\!=\!\int|\chi(z)|^4dz$. (This value certainly is not equal to
the quantum well width $d_{\rm QW}$, but constitutes a fraction of
the latter, e.g.: $d/d_{\rm QW}\simeq 1/3$.) Formula
\eqref{hftime} has been obtained for the case of unpolarized
nuclei, i.e. $\overline{M_n}\!=\!0$, and the correlation length of
the nuclear momenta distribution is smaller than the magnetic
length $l_B$, hence $\overline{M_n^2}\!=\!5/4$ where the
over-line means averaging over the volume $2\pi l_B^2d$.

The elementary process just studied characterizes only the initial
stage of the Goldstone condensate breakdown. Further physical
picture of the relaxation follows absolutely the same scenario
that was analyzed in Ref. [\onlinecite{di04}]. When the Goldstone
condensate is depleted, a ``thermodynamic condensate'' is
developing. The latter is formed by spin waves with nonzero but
negligibly small wave vectors, which are of the order of or
smaller than the uncertainty value determined by smooth disorder.
The number of nonzero excitons is equal to $|\Delta S|$, i.e. to
deviation of the QHF total spin number $S$ from its ground state
value. $|\Delta S(t)|$ reaches the maximum value [still being well
smaller than the simultaneous number $N(t)$], and in the final
stage both condensates decay. By
considering concentrations of the Goldstone and thermodynamic
condensates -- $\;n\!=\!N/\nu\,{}'\!N_\phi$ and $m\!=\!|\Delta
S|/\nu\,{}'\!N_\phi$, one can find equations governing the
relaxation, \vskip -6mm
\begin{equation}\label{releq}
\tau_{\rm hf}dn/dt=-2n^2-4mn\quad \mbox{and}\quad \tau_{\rm
hf}dm/dt=n^2-2m^2\, .
\end{equation}
These equations yield $n(t)\!=\!1/{[2n(0)t^2\!+\!2t\!+\!1/n(0)]}$
and $m=n(t)n(0)t$, where $t$ is measured in $\tau_{\rm hf}$. I
remind that, as in the work of  [\onlinecite{di04}], this result
is analytical but still approximate -- $\;$it should work well if
$n(0)\!<\!1/2$.

The only difference is thereby a change in Eqs. \eqref{releq} from
the characteristic relaxation time [\onlinecite{di04}] \vskip
-10mm
\begin{equation}\label{SOtime}
1/\tau_{\rm so}={8\pi^2(\alpha^2\!+\!\beta^2)M_{\mbox{\scriptsize  x}}^2\epsilon_Z
  {\overline K}(q^*)}/{\hbar^3\omega_c^2 l_B^4}\,,
\end{equation}
determined by the SO coupling and smooth random potential, to the
HF coupling time $\tau_{\rm hf}$ \eqref{hftime}. (I keep notations
of the paper \cite{di04}: $\alpha$ and $\beta$ are the Rashba and
Dresselhaus SO parameters, ${\overline K}(q)$ stands for the
Fourier component of the smooth random potential correlator.)
Comparing $\tau_{\rm so}$ with $\tau_{\rm hf}$, one can note that
they have opposite dependences on the magnetic field. If $K(r)$ is
Gaussian, then ${\overline K}(q)$ is sharply decreasing with $B$:
${\overline
K}(q^*)\!=\!(\Delta^2\Lambda^2/4\pi)\exp{\!\left(-M_{\rm
x}\epsilon_Z\!\Lambda^2/2l_B^2\right)}\sim\exp{\!(-\gamma M_{\rm
x}B^2)}$. ($\Delta$ and $\Lambda$ stand for the amplitude and
correlation length of the random potential, respectively.)
Meanwhile, according to the above calculations, the HP rate
$1/\tau_{\rm hp}$ is proportional to the squared local density
$|\Psi(\mbox{\boldmath $R$}_n)|^4\!\sim\! B^2$ and to the number
of nuclei per  electron $2\pi l_B^2d/v_0\!\sim \!1/B$; therefore
$1/\tau_{\rm hp}\!\sim \! M_{\rm x}B$.

More specific estimates of $\tau_{\rm hp}$ and $\tau_{\rm so}$ are
required for appropriate comparison of both relaxation channels.
The material parameters and characteristic parameters related to
modern wide quantum-well structures could be, e.g., chosen as
$\;v_0\!\left(A_{\rm Ga}^2\!+\!A_{\rm
As}^2\right)\!\!=\!\!1.8\!\cdot\!10^{-4}\,$meV${}^2$nm${}^3$,
$2(\alpha^2\!+\!\beta^2)/(l_B \hbar\omega_c)^2\!\!=
\!\!10^{-3}/B$, $\epsilon_Z\!=\!\!0.0255B\,$meV,
$\Lambda\!=\!60\,$nm, $\Delta\!=\!0.3\,$meV, and $d\!=\!8\,$nm
(here $B$ is assumed to be measured in Tesla; c.f. also estimates
in Ref. \cite{di09}). However, estimate of the effective
spin-exciton mass $M_{\rm x}$ strongly depends on the
finite thickness formfactor. There are experimental data
where $M_{\rm x}$ is found at comparatively low magnetic fields:
(i) $1/M_{\rm x}\!\approx\!1.2\,$meV at $B\!=\!2.27\,$T and
$\nu\!=\!1$ in the $33\,$nm quantum well \cite{ga08}; (ii)
$1/M_{\rm x}\!\approx\!1.51\,$meV at $B\!=\!2.69\,$T and
$\nu\!=\!1$ in the $23\,$nm quantum well \cite{kukush09}; and
(iii) $1/M_{\rm x}\!\approx\!0.44\,$meV at $B\!=\!2.9\,$T and
$\nu\!=\!1/3$ in the $25\,$nm quantum well \cite{kukush06}. For
these fields characterized by the inequality $l_B\!>\!d$, the
$B$-dependence should be approximately $1/M_{\rm x}\sim B^{1/2}$,
but in the $l_B\!<\!d$ strong field regime the inverse mass grows
much weaker with $B$. Based on these data, the semi-empirical
analysis using characteristic GaAs/AlGaAs formfactors allows me to
consider values $1/M_{\rm x}\!\simeq 2\,$meV at $\nu\!=\!1$ and
$1/M_{\rm x}\!\simeq 0.67\,$meV at $\nu\!=\!1/3$ as the
characteristic ones for the $10\,$T$<\!B\!<\!20\,$T range. (Note
that at a given field $B$ the estimate $M_{\rm
x}^{-1}|_{\nu\!<\!1}\!\simeq\!\nu{\,}'\!\cdot\!M_{\rm
x}^{-1}|_{\nu\!=\!1}$ holds according to the semi-phenomenological
theory \cite{lo93}.) Then, if substituting the above parameters
into Eqs. \eqref{hftime} and \eqref{SOtime}, one obtains
$\tau_{\rm hf}(B^*)\!=\!\tau_{\rm so}(B^*)$ at
$B^*\!\approx\!15\,$T if $\nu\!=\!1$, or at $B^*\!\approx\!9.3\,$T
if $\nu\!=\!1/3$. The characteristic relaxation time at these
crossover points constitutes $\simeq 4\,\mu$s or $\simeq 2\,\mu$s
respectively.

To conclude, I have reported on a new spin relaxation channel in
the spin polarized strongly correlated 2DEG. The mechanism
involves only the hyperfine coupling to GaAs nuclei, and no other
interactions are required for this relaxation channel to be
realized. The problem is solved by using the {\it excitonic
representation} technique. Although the Goldstone mode relaxation
in a QHF occurs by the scenario studied earlier \cite{di04}, a
crossover from the SO characteristic relaxation time
\eqref{SOtime} to the hyperfine coupling time \eqref{hftime}
occurs in a strong magnetic field $B\gapprox 10\,$T.

The work is supported by the RFBR.


\begin{thebibliography}{99}
\bibitem{foot0} {Producing the GM in a QHF
 seems  to be a difficult experimental task, and it is absolutely clear that authors of recent
 experiments [\onlinecite{fu08}] still studied the spin relaxation
 for the initial deviation of the first type. In this connection,
 as far as the authors refer to my publication [\onlinecite{di04}]
 devoted to the GM relaxation, I note that the reference to the
 work [\onlinecite{di09}] would be much more appropriate.}
\bibitem{foot1} {The diagonal contribution of the hyperfine coupling Hamiltonian is negligible.
Besides, I ignore the energy of nuclei which consists of the
contribution due to their interaction extraneous to electrons and
of the nuclear Zeeman energy. Variations of both, associated with
the change of nuclei spins, are negligibly small due to the tiny
nuclear magnetic momentum.}
\end{thebibliography}
\end{document}